\documentclass[aps,prb,amsmath,amssymb,reprint,superscriptaddress]{revtex4-1}

\usepackage{graphicx}
\usepackage{epstopdf}
\usepackage{natbib}
\begin{document}

%\preprint{}

\title{General Quantum Fidelity Susceptibilities for the $J_1-J_2$ Chain}


\author{Mischa Thesberg}
\email[]{thesbeme@mcmaster.ca}
%\thanks{}
\author{Erik S. S{\o}rensen}
\email[]{sorensen@mcmaster.ca}
\homepage[]{http://comp-phys.mcmaster.ca}
 \affiliation{Department of Physics \& Astronomy, McMaster University\\1280 Main St. W., Hamilton ON L8S 4M1, Canada.}


\date{\today}

\begin{abstract}
We study slightly generalized quantum fidelity susceptibilities where the
differential change in the fidelity is measured with respect to a different
term than the one used for driving the system towards a quantum phase transition.
As a model system we use the spin-$1/2$ $J_1-J_2$ antiferromagnetic Heisenberg chain.
For this model, we study three fidelity susceptibilities,
$\chi_{\rho}$, $\chi_D$ and $\chi_\mathrm{AF}$, which are related to the spin stiffness,
the dimer order and antiferromagnetic order, respectively.
All these ground-state fidelity susceptibilities are sensitive to the phase diagram
of the $J_1-J_2$ model. We show that they all can accurately identify a quantum critical
point in this model occurring at $J_2^c\sim0.241 J_1$ between a gapless Heisenberg phase for
$J_2<J_2^c$ and a dimerized phase for $J_2>J_2^c$. This phase transition, in the Berezinskii-Kosterlitz-Thouless universality class, is controlled by
a marginal operator and is therefore particularly difficult to observe.
\end{abstract}

%\keywords{}

\maketitle

\section{Introduction}

The study of quantum phase transitions, especially in one and two dimensions,
is a topic of considerable and ongoing interest.\cite{Sachdev_QPT_Book}
Recently the utility of a concept with its origin in quantum information, the
quantum fidelity and the related fidelity susceptibility, was demonstrated for
the study of quantum phase transitions (QPT).\cite{Quan_Original,Zanardi_Original,Zhou_Original,You}  It
has since then been successfully applied to a great number of
systems.\cite{Zanardi_Free_Fermion_Systems,Cozzini_Graphs,Cozzini_Matrix_Product_States,Buonsante_Bose_Hubbard,Schwandt:2009,Albuquerque:2010}
In particular, it has been applied to the $J_1-J_2$ model that we consider here.~\cite{Chen_J1_J2}.
For a recent review of the fidelity approach to quantum phase transitions, see Ref.~\onlinecite{Gu_Review}.
Most of these studies consider the case where the system undergoes a quantum phase transition as
a coupling $\lambda$ is varied. The quantum fidelity and fidelity susceptibility is then defined
with respect to the same parameter. Apart from a few studies,~\cite{Zanardi:2007,Venuti_Scaling,DeGrandi:2011} relatively little attention has been given to the case
where the quantum fidelity and susceptibility are defined with respect to a coupling different than
$\lambda$. Here we consider this case in detail for the $J_1-J_2$ model and show that, if appropriately defined,
these general fidelity susceptibilities may yield considerable information about quantum phase transitions
occurring in the system and can be very useful in probing for a non-zero order parameter. 

Without loss of generality, the Hamiltonian of any many-body system can be written as
\begin{eqnarray}
\label{General_Hamiltonian}
H(\lambda) = H_0 + \lambda H_\lambda, 
\end{eqnarray} 
where $\lambda$ is a variable which typically parametrizes an
interaction and exhibits a phase transition at some critical value $\lambda_c$.  
In this form $H_\lambda$ is then recognized as a term that \emph{drives} the phase transition.\cite{You} 
Using the eigenvectors of this Hamiltonian the ground-state (differential) fidelity can then be written as:
\begin{eqnarray}
F(\lambda)=\left| \langle \Psi_0 (\lambda) \vert \Psi_0 (\lambda + \delta \lambda) \rangle \right|.
\end{eqnarray}
A series expansion of the GS fidelity in $\delta\lambda$ yields
\begin{eqnarray}
\label{F_Definition}
F(\lambda)=1-\frac{(\delta\lambda)^2}{2}\frac{\partial^2 F}{\partial \lambda^2} + \ldots
\end{eqnarray}
where $\partial_{\lambda}^2 F\equiv\chi_{\lambda}$ is called the
\emph{fidelity susceptibility}. For a discussion of sign conventions
    and a more complete derivation see the topical review
    by Gu, Ref.~\onlinecite{Gu_Review}.  If the higher-order terms are
taken to be negligibly small then the fidelity susceptibility
is defined as:
\begin{eqnarray}
\label{FS_Definition}
\chi_{\lambda} (\lambda) = \frac{2(1-F(\lambda))}{(\delta\lambda)^2}
\end{eqnarray}
The scaling of $\chi_\lambda$ at a quantum critical point, $\lambda_c$, is often of considerable interest and
has been studied in detail and previous studies~\cite{Zanardi:2007,Venuti_Scaling,Gu:2009,Schwandt:2009,Albuquerque:2010}
have shown that
\begin{equation}
\chi_\lambda \sim L^{2/\nu}, \ \ \ \ \chi_\lambda/N\sim L^{2/\nu-d},
\label{eq:chi_scaling}
\end{equation}
with $N=L^d$ the number of sites in the system.
An easy way to re-derive this result is by envoking finite-size scaling. Since $1-F$ obviously is {\it dimensionless} it
follows from Eq.~(\ref{FS_Definition}) that the appropriate finite-size scaling form for $\chi_\lambda$ is
\begin{equation}
\label{eq:finscaling}
\chi_\lambda \sim (\delta\lambda)^{-2}f(L/\xi).
\end{equation}
If we now consider the case where 
the parameter $\lambda$ drives the transition we may at the critical point $\lambda_c$ identify $\delta\lambda$ with $\lambda-\lambda_c$. 
It follows that  $\xi\sim(\delta\lambda)^{-\nu}$. 
As usual, we can then replace $f(L/\xi)$ by an equivalent function $\tilde f(L^{1/\nu}\delta\lambda)$. The requirement that $\chi_\lambda$ remains finite for a finite system
when $\delta\lambda\to 0$ then implies that to leading order $\tilde f(x)\sim x^2\sim L^{2/\nu}(\delta\lambda)^2$, from which Eq.~(\ref{eq:chi_scaling}) follows. 

Here we shall consider a slightly more general case where the term driving the quantum phase transition
is not the same as the one with respect to which the fidelity and fidelity susceptibility are defined.
That is, one considers:
\begin{eqnarray}
\label{GH2}
H(\lambda,\delta) = H_1 + \delta H_I, \ \ H_1=H_0 + \lambda H_\lambda.
\end{eqnarray} 
The fidelity and the related susceptibility is then defined as
\begin{eqnarray}
F(\lambda,\delta)&=&\left| \langle \Psi_0 (\lambda,0) \vert \Psi_0 (\lambda,\delta) \rangle \right|, \\
\chi_{\delta} (\lambda)& = &\frac{2(1-F(\lambda,\delta))}{\delta^2}
\end{eqnarray}
The scaling of $\chi_{\delta}$ at $\lambda_c$ for this more general case was
derived by Venuti \emph{et al.}\cite{Venuti_Scaling} where it was shown that:
\begin{eqnarray}
\label{Chi_Scaling}
\chi_\delta \sim L^{2d+2z-2\Delta_v}, \ \ \ \chi_\delta/N\sim L^{d+2z-2\Delta_v}.
\end{eqnarray}
Here, $z$ is the dynamical exponent, $d$ the dimensionality and $\Delta_v$ 
the scaling dimension of the perturbation $H_I$.  In all cases that we consider here $z=d=1$. 
We note that
Eq.~(\ref{Chi_Scaling}) assumes $[H_1,H_I]\neq 0$, if $H_I$ commutes with $H_1$
then $F=1$ and $\chi_\delta=0$.  The case where $H_\lambda$ and $H_I$ coincide
is a special case of Eq.~(\ref{Chi_Scaling}) for which $\Delta_V=d+z-1/\nu$ and
one recovers Eq.~(\ref{eq:chi_scaling}).

A particular appealing feature of Eq.~(\ref{eq:chi_scaling}) is that when $2/\nu>d$,
$\chi_\lambda/N$ will diverge at $\lambda_c$ and the fidelity susceptibility can
then be used to locate the $\lambda_c$ {\it without} any need for knowing the order
parameter.  Secondly, it can be shown~\cite{You,Zanardi:2007} that the fidelity
susceptibility can be expressed as the zero-frequency {\it derivative} of the
dynamical correlation function of $H_I$, making it a very
sensitive probe of the quantum phase transition.~\cite{Chen_Perturbation} On the other hand, if a phase
transition is expected one might then use the fidelity susceptibility as a very
sensitive probe of the order parameter through a suitably defined $H_\delta$ in
Eq.~(\ref{GH2}). This is the approach we shall take here using the $J_1-J_2$
spin chain as our model system. 

The spin-$1/2$ Heisenberg $J_1-J_2$ chain is a very well studied model.
The Hamiltonian is:
\begin{eqnarray}
\label{Hamiltonian}
H =  \sum_i S_i \cdot S_{i+1} + J_2 \sum_i S_i \cdot S_{i+2}
\end{eqnarray}
where  $J_2$ is understood to be the
ratio of the next-nearest neighbor exchange parameter over the nearest
neighbor exchange parameter ($J_2=J_2'/J_1'$).  This model is known to have
a quantum phase transition of the  Berezinskii-Kosterlitz-Thouless (BKT) universality class occurring at $J_2^c$ between a gapless 
'Heisenberg' (Luttinger liquid) phase for $J_2<J_2^c$ and a dimerized phase with a two-fold
degenerate ground-state for $J_2>J_2^c$.
Field theory\cite{J1_J2_Field_Theory_1,Eggert:1992}, exact
diagonalization\cite{J1_J2_ED_1,Eggert} and DMRG\cite{DMRG_2,DMRG_3}, have
yielded very accurate estimates of the Luttinger Liquid-Dimer phase transition,
the most accurate of these being due to Eggert\cite{Eggert} which yielded a value of
$J_2^c=0.241167$.   
Previous studies by Chen \emph{et al.}~\cite{Chen_J1_J2} of this model using the fidelity approach
used the same term for the driving and perturbing part of the Hamiltonian
as in Eq.~(\ref{General_Hamiltonian}) with the correspondence $H_0=\sum_i
S_i \cdot S_{i+1}$, $H_\lambda=\sum_i S_i \cdot S_{i+2}$, $\lambda=J_2$.\cite{Chen_J1_J2} . 
Chen \emph{et al.} demonstrated that, though no useful
information about the Luttinger Liquid-Dimer phase transition could be obtained
directly from the {\it ground-state} fidelity (and similarly the fidelity
susceptibility), a clear signature of the phase transition was present in the fidelity
of the {\it first excited} state.\cite{Chen_J1_J2} Sometimes this is taken as an indication
that ground-state fidelity susceptibilities are not useful for locating a quantum phase transition
in the BKT universality class. Here we show that more general ground-state fidelity susceptibilities indeed
can locate this transition.

Specifically, we will study three fidelity susceptibilities, 
$\chi_{\rho}$, $\chi_D$ and $\chi_\mathrm{AF}$, which are coupled to the spin stiffness,
a staggered interaction term and a staggered field term, respectively.  In section~\ref{sec:chirho} we present our results
for $\chi_\rho$ while section~\ref{sec:chiD} is focused on $\chi_D$ and section~\ref{sec:chiAF} on $\chi_\mathrm{AF}$.

\section{The Spin Stiffness Fidelity Susceptibility, $\chi_{\rho}$}\label{sec:chirho}
We begin by considering the $J_1-J_2$ model with $J_2=0$ but with an anisotropy term $\Delta$, what is
usually called the $XXZ$ model:
\begin{eqnarray}
H_{\mathrm{XXZ}}=\sum_{i}[\Delta S_i^z S_{i+1}^z + \cfrac{1}{2}(S_i^+ S_{i+1}^- + S_i^- S_{i+1}^+)].
\end{eqnarray}
The Heisenberg phase of this model, occurring for $\Delta\in [-1,1]$, is
characterized by a non-zero spin stiffness~\cite{Shastry:1990,Sutherland:1990}
defined as:
\begin{equation}
\rho(L) = \left. \frac{\partial^2 e(\phi)}{\partial \delta^2} \right|_{\phi=0}.
\end{equation}
Here, $e(\phi)$ is the ground-state energy per spin of the model where a twist of $\phi$ is applied at every bond:
\begin{eqnarray}
&&H_{\mathrm{XXZ}}(\Delta,\phi)=\nonumber\\
    &&\sum_{i}[\Delta S_i^z S_{i+1}^z + \cfrac{1}{2}(S_i^+ S_{i+1}^-e^{i\phi} + S_i^- S_{i+1}^+e^{-i\phi})].
    \label{eq:hxp}
\end{eqnarray}
The spin stiffness can be calculated exactly for the $\mathrm{XXZ}$ model for finite $L$ using the Bethe ansatz,~\cite{Sorensen_XXZ}
and exact expressions in the thermodynamic limit are available. ~\cite{Shastry:1990,Sutherland:1990}
Interestingly the usual fidelity susceptibility with respect to $\Delta$ can also be calculated exactly.~\cite{Yang:2007,Fjaerestad:2008}

\begin{figure}[t]
\includegraphics[scale=0.35]{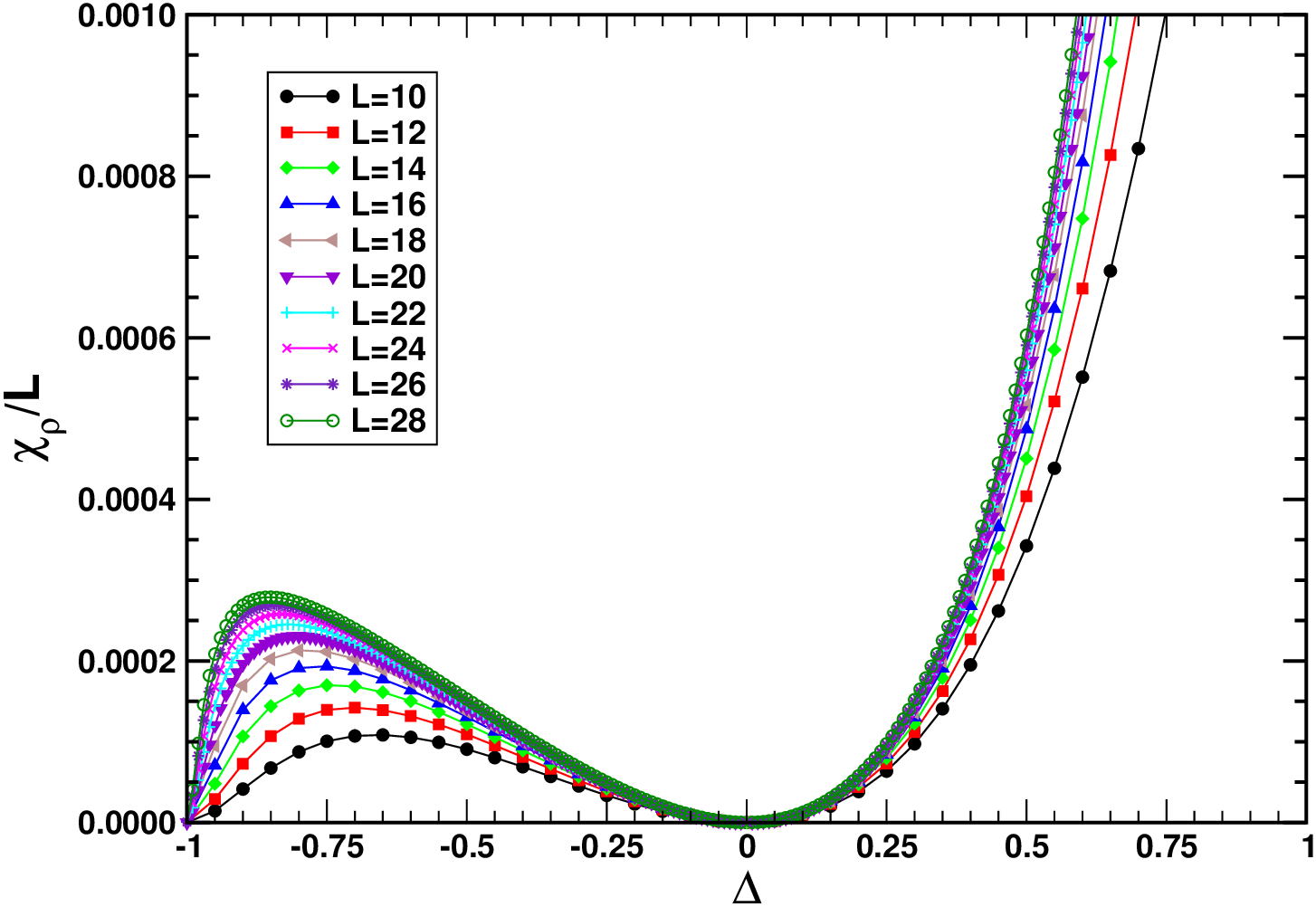}
\caption{\label{Chi_Rho_vs_XXZ_Delta}  (Color online.) $\frac{\chi_{\rho}}{L}$ vs. $\Delta$:
  The spin stiffness fidelity susceptibility ($\chi_{\rho}(\Delta)/L$) as a
    function of the z-anistropy $\Delta$.  At the $\Delta=0$ point the
    spin-current operator $\mathcal{J}$ and kinetic energy $\mathcal{T}$ commute with the XXZ Hamiltonian and thus
    such a perturbation does not change the ground-state, and the fidelity is
    one.  Thus, $\chi_{\rho}$ is zero at this point. } \end{figure}

Since the non-zero spin stiffness defines the gapless Heisenberg phase it is therefore of interest to define
a fidelity susceptibility associated with the stiffness. This can be done through the overlap of the ground-state with
$\phi=0$ and a non-zero $\phi$. 
With $\Psi_0(\Delta,\phi)$ the ground-state
of $H_\mathrm{XXZ}(\Delta,\phi)$ we can define the fidelity and fidelity susceptibility with respect to the twist in the limit
where $\phi\to 0$:
\begin{eqnarray}
F(\Delta,\phi)&=&\left| \langle \Psi_0 (\Delta,0) \vert \Psi_0 (\Delta , \phi) \rangle \right|,\\
\chi_{\rho} (\Delta)& = &\frac{2(1-F(\Delta,\phi))}{\phi^2}.\label{eq:chirho}
\end{eqnarray}
To calculate $\chi_{\rho}$ the ground-state of the unperturbed Hamiltonian was
calculated through numerical exact diagonalization.  The system was then
perturbed by adding a twist of $e^{i \phi}$ at each bond and recalculating the ground-state.
From the corresponding fidelity, $\chi_{\rho}$ was calculated using Eq.~(\ref{eq:chirho}). 
Our results for $\chi_\rho/L$ versus $\Delta$ are shown in Fig.~\ref{Chi_Rho_vs_XXZ_Delta}.
For all data $\phi$ was taken to be $10^{-3}$ and periodic boundary
conditions were assumed.  In all cases it was verified that the finite value of $\phi$ used had
no effect on the final results. The numerical diagonalizations were done using the
Lanczos method as outlined by Lin \emph{et al.} \cite{Lin}  Total $S^z$
symmetry and parallel programming techniques were employed to make computations
feasible. Numerical errors are small and conservatively estimated to be on the
order of $10^{-10}$ in the computed ground-state energies.  

In order to understand the results in Fig.~\ref{Chi_Rho_vs_XXZ_Delta} in
more detail we expand Eq.~(\ref{eq:hxp}) for small $\phi$:
\begin{eqnarray}
&&H_{\mathrm{XXZ}}(\Delta,\phi)\sim H_{\mathrm{XXZ}}(\Delta)+\phi\mathcal{J}-\frac{\phi^2}{2}\mathcal{T}+\ldots,\label{eq2}\\
&&\mathcal{J}=    \frac{i}{2}\sum_{i}(S_i^+ S_{i+1}^-- S_i^- S_{i+1}^+),\label{eq:J}\\
&&\mathcal{T}=    \frac{1}{2}\sum_{i}(S_i^+ S_{i+1}^-+ S_i^- S_{i+1}^+).
    \label{eq:hxp2}
\end{eqnarray}
Here, $\mathcal{J}$ is the spin current and $\mathcal{T}$ a kinetic energy term.
The first thing we note is that, when $\Delta=0$ both $\mathcal{J}$ and $\mathcal{T}$
commute with $H_{\mathrm{XXZ}}(\Delta=0)$. The ground-state wave-function is therefore
independent of $\phi$ (for small $\phi$) and $\chi_\rho\equiv 0$. This can clearly be seen
in Fig.~\ref{Chi_Rho_vs_XXZ_Delta}.
\begin{figure}[t]
\includegraphics[scale=0.35]{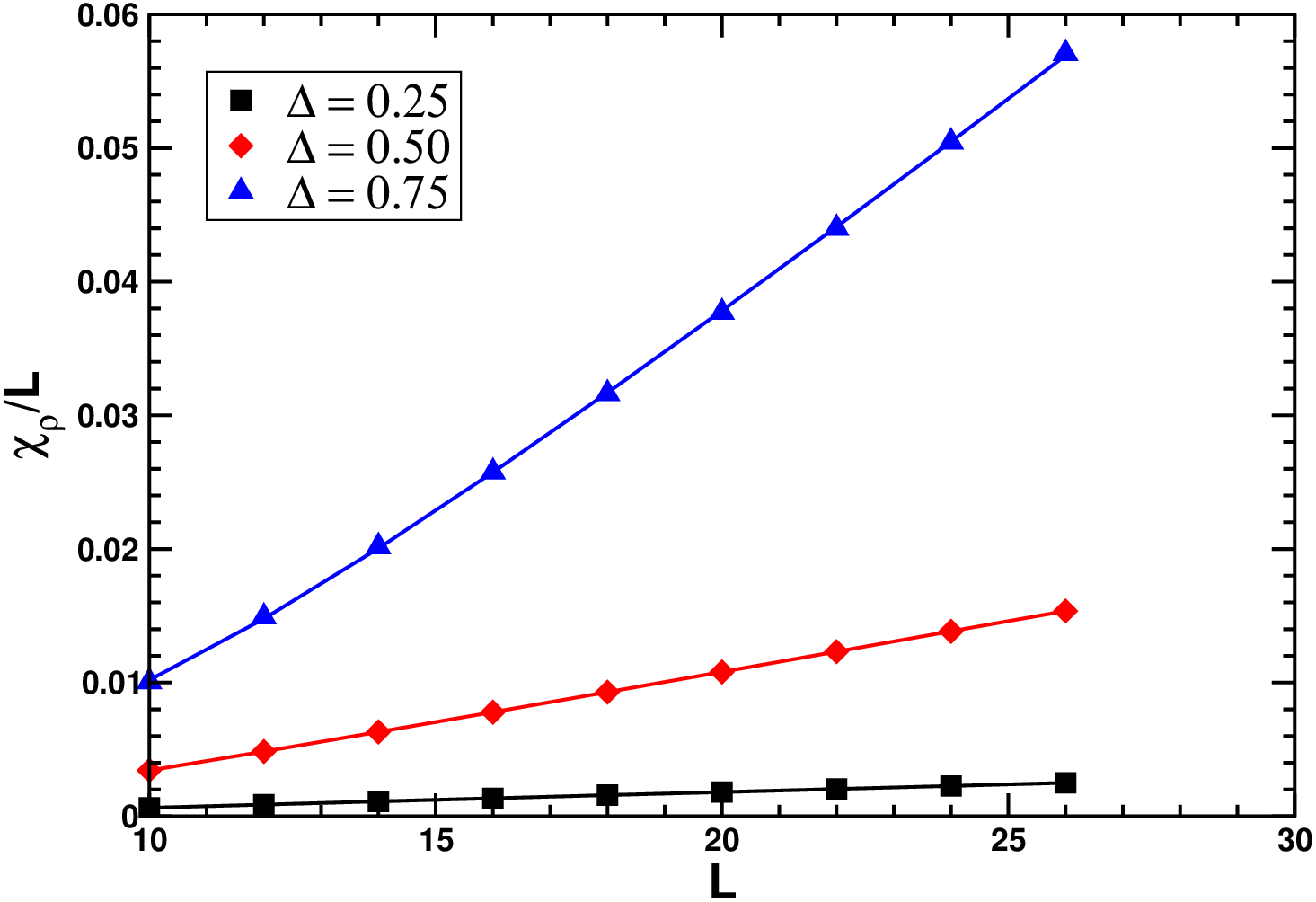}
\caption{\label{Chi_Rho_Scaling}(Color online.)   $\chi_{\rho}$ vs. $L$ (the XXZ model at different
    values of $\Delta$): This graph shows the scaling of  $\chi_{\rho}$ with
  system size for different values of the z-anisotropy $\Delta$.  The points
  represent numerical data and the lines represent fits to the scaling form predicted for the
  spin stiffness susceptibility $\chi_\rho/L=A_1L+A_2+A_3L^{-1}+A_4L^{3-8K}$.
It can be seen that there is good agreement.} \end{figure}

In the continuum limit the spin current $\mathcal{J}$ can be expressed in an effective low energy field
theory~\cite{Sirker:2011} with scaling dimension $\Delta_\mathcal{J}=1$. However, we expect
subleading corrections to arise from the presence of the operators $(\partial_x\Phi)^2$ with
scaling dimension 2 and $\cos(\sqrt{16\pi K}\Phi)$ with scaling dimension $4K$. Here, $K$ is given
by $K=\pi/(2(\pi-\mathrm{arcos}(\Delta)))$. For $\Delta\neq 0$ both of these terms will be generated by the
term $\mathcal{T}$ in Eq.~(\ref{eq2}).~\cite{Venuti_Scaling}
With these scaling dimensions and with the use of Eq.~(\ref{Chi_Scaling}) we then find:
\begin{equation}
\chi_\rho/L=A_1L+A_2+A_3L^{-1}+A_4L^{3-8K}
\end{equation}
In Fig.~\ref{Chi_Rho_Scaling} a fit to this form is shown for 3 different values of $\Delta=0.25,0.5$ and
$0.75$ in all cases do we observe an excellent agreement with the expected form with corrections arising
from the last term $L^{3-8K}$ being almost un-noticeable until $\Delta$ approaches 1. We would expect
the sub-leading corrections $L^{-1}$ and $L^{3-8K}$ to be absent if the perturbative term is just
$\phi\mathcal{J}$.

We now turn to a discussion of a definition of $\chi_\rho$ in the presence of a non-zero $J_2$ but
restricting the discussion to the isotropic case $\Delta=1$.
In this case we define:
\begin{eqnarray}
&&H(\phi)=\nonumber\\
    &&\sum_{i}[S_i^z S_{i+1}^z + \cfrac{1}{2}(S_i^+ S_{i+1}^-e^{i\phi} + S_i^- S_{i+1}^+e^{-i\phi})]+\nonumber\\
    &&J_2\sum_{i}[S_i^z S_{i+2}^z + \cfrac{1}{2}(S_i^+ S_{i+2}^-e^{i\phi} + S_i^- S_{i+2}^+e^{-i\phi})].
    \label{eq:hj2}
\end{eqnarray}
That is, we simply apply the twist $\phi$ at every bond of the Hamiltonian.
As before we can expand:
\begin{eqnarray}
&&H(\phi)\sim H(0)+\phi(\mathcal{J}_1+\mathcal{J}_2)-\frac{\phi^2}{2}(\mathcal{T}_1+\mathcal{T}_2)+\ldots,\label{eq3}\\
&&\mathcal{J}_1=    \frac{i}{2}\sum_{i}(S_i^+ S_{i+1}^-- S_i^- S_{i+1}^+),\label{eq:J1}\\
&&\mathcal{J}_2=    \frac{i}{2}\sum_{i}(S_i^+ S_{i+2}^-- S_i^- S_{i+2}^+),\label{eq:J2}\\
&&\mathcal{T}_1=    \frac{1}{2}\sum_{i}(S_i^+ S_{i+1}^-+ S_i^- S_{i+1}^+),\\
&&\mathcal{T}_2=    \frac{1}{2}\sum_{i}(S_i^+ S_{i+2}^-+ S_i^- S_{i+2}^+).
    \label{eq:hxp3}
\end{eqnarray}

\begin{figure}[t]
\includegraphics[scale=0.35]{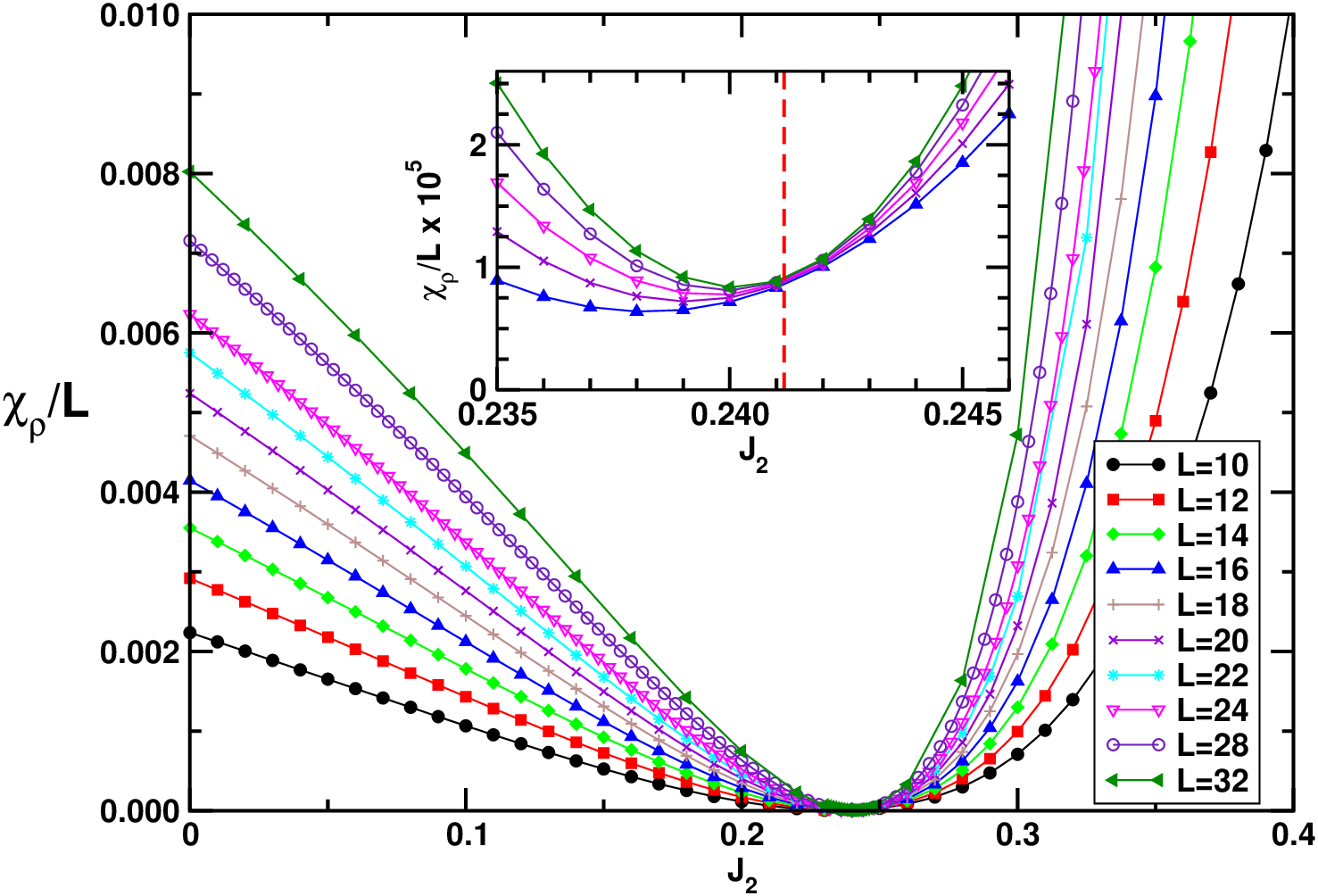}
\caption{\label{Chi_Rho_vs_J2_Plus_Inset} (Color online.) $\frac{\chi_{\rho}}{L}$ vs. $J_2$ and
  Inset: The generalized spin stiffness susceptibility, $\chi_{\rho}$ as a
    function of the second nearest-neighbor exchange coupling $J_2$.  The
    system acquires a clearly size invariant form in the vicinity of the
    critical point $J_2 \sim 0.24$ (as well as tending to a global minima).
    Inset shows the the minima for system sizes L=16,20,24,28,32 with $J_2^c$ indicated as the
    vertical dashed line. A clear dependence of the $J_2$ value of
    $\chi_{\rho}/L$ minima on the system size can be seen. } \end{figure}
\begin{figure}[t]
\includegraphics[scale=0.35]{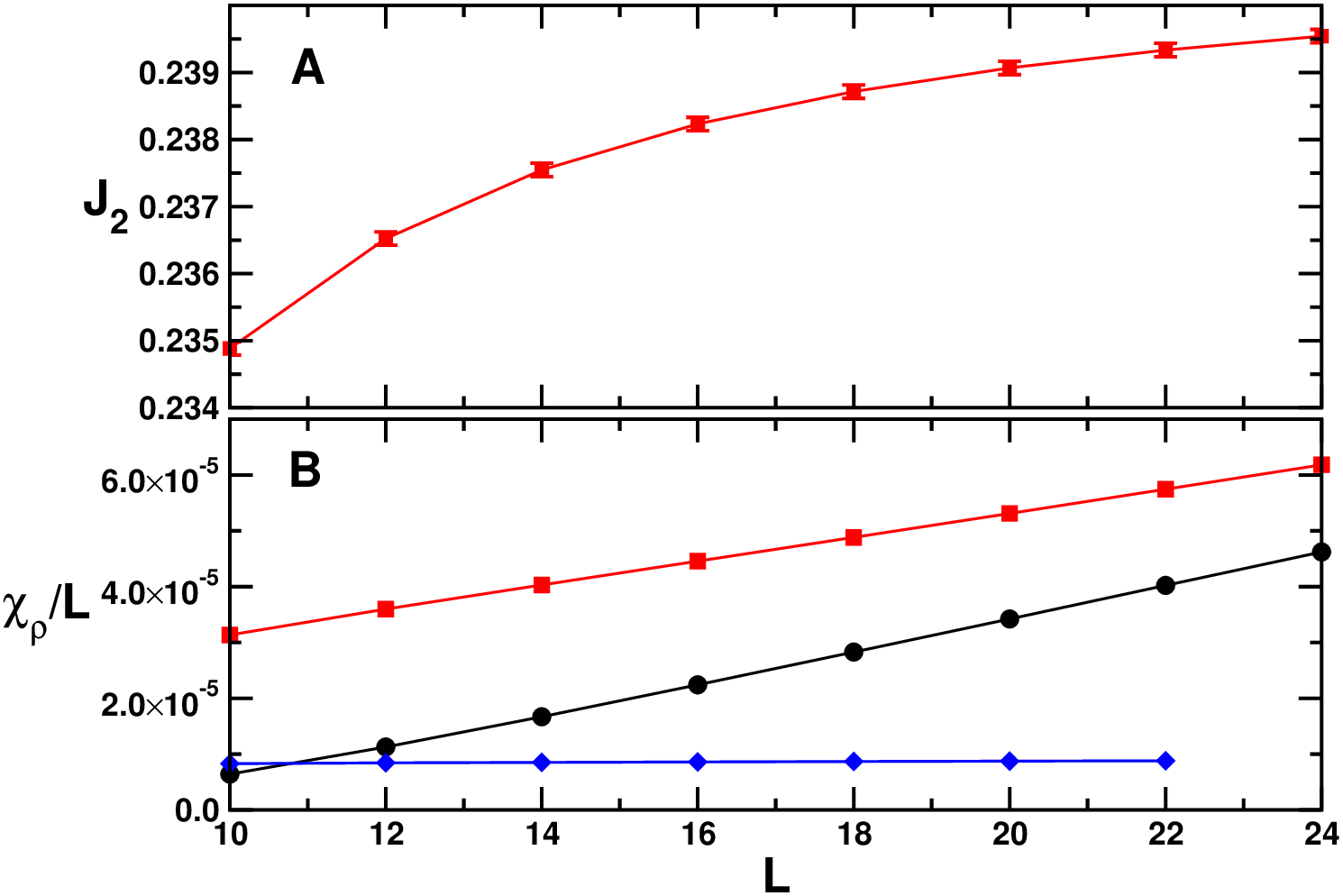}
\caption{\label{Chi_Rho_Scaling_and_Convergence} (Color online.)  
  A: The $J_2$ value of $\chi_{\rho}$ minima as a function of system
  size, as well as a (power law) line of best fit.  As the system size tends
  towards infinity the power law best fit predicts a minima at $J_2=0.24077$ in
  good agreement with previously published results.  B: Scaling of
  $\chi_{\rho}$ at $J_2=0.23$ (the highest, linear curve), $J_2=0.25$ (the
      second highest, linear curve) and at the critical point $J_2=0.241167$
  (flat curve).  The near constant scaling of $\frac{\chi_{\rho}}{L}$ at the
  critical point as well as non-constant scaling on either side of the critical
  point can clearly be seen. } 
\end{figure}

Our results for $\chi_\rho/L$ versus $J_2$ using this definition are shown in Fig.~\ref{Chi_Rho_vs_J2_Plus_Inset} for
a range of $L$ from 10 to 32.
In the region of the critical point at $J_2=0.241167$ the size
dependence  of $\chi_\rho/L$ vanishes yielding near scale invariance. 
How well this works close to $J_2^c$ is shown in the inset of Fig.~\ref{Chi_Rho_vs_J2_Plus_Inset}.
This alone can be taken to
be a strong indication of $\chi_{\rho}/L$s sensitivity to the phase transition.
In fact, this scale invariance works so well that one can locate the critical
point to a high precision simply by verifying the scale invariance. This is illustrated
in Fig.~\ref{Chi_Rho_Scaling_and_Convergence}B where $\chi_\rho/L$ is plotted as a function
of $L$ for $J_2=0.23$, $J_2=J_2^c$ and $J_2=0.25$. From the results in Fig.~\ref{Chi_Rho_Scaling_and_Convergence}B
the critical point $J_2^c$ where $\chi_\rho/L$ becomes independent of $L$ is immediately
visible.

As can be seen in the inset of Fig.~\ref{Chi_Rho_vs_J2_Plus_Inset} $\chi_\rho/L$ reaches a minimum
slightly prior to $J_2^c$.
The $J_2$ value at which this minimum occurs has
a clear system size dependence which can be fitted to a power-law and
extrapolated to $L=\infty$ yielding a value of $J_{2c}=0.24077$. Hence,
the minimum coincides with $J_2^c$ in the thermodynamic limit. This
is shown in Fig.~\ref{Chi_Rho_Scaling_and_Convergence}A.
Comparison of this value with the accepted
$J_2^c=0.241167$ reveals impressive agreement.  Another noteworthy feature of the
results in Fig.~\ref{Chi_Rho_vs_J2_Plus_Inset} is that $\chi_\rho/L$ is {\it non-zero}
at the critical point, $J_2^c$. This value is very small but we have verified in detail
that numerically it is non-zero. 

The scale invariance of $\chi_\rho/L$ is clearly induced by the disappearance~\cite{Eggert:1992}
of the marginal operator $\cos(\sqrt{16\pi K}\Phi)$ at $J_2^c$. We expect that in the continuum limit 
the absence of this operator
implies that the spin current commutes with the Hamiltonian resulting in $\chi_\rho$ being effectively
zero at $J_2^c$. The observed non-zero value of $\chi_\rho/L$ would then arise from short-distance physics.

Note that, as mentioned previously, we take the spin
stiffness to be represented by a twist on \emph{every} bond, both first and
second nearest neighbor and not merely on the boundary as is sometimes done.
This choice is not just a matter of taste.  Imposing a twist only on the
boundary (usually) breaks the translational invariance of the ground-state and, through
extension, effects the value and behavior of the fidelity itself.  Another
point of note is the use of a twist of only $\phi$ between
next-nearest neighbors.  Geometric intuition would suggest that a twist of
$2\phi$ should be applied between next-nearest neighbor bonds.
However, for
the small system sizes available for exact diagonlization it is found that a simple twist of $\phi$ on
both bonds yields {\it significantly} better scaling.

\section{The Dimer Fidelity Susceptibility, $\chi_D$}\label{sec:chiD}

We now turn to a discussion of a fidelity susceptibility associated with the
dimer order present in the $J_1-J_2$ model for $J_2>J_2^c$.
This susceptibility, which we call $\chi_D$, is coupled to the order parameter of
the dimerized phase by design.  
Usually in the fidelity approach to quantum phase transitions one considers the
case where the ground-state is unique in the absence of the perturbation. This
is not the case here, leading to a diverging $\chi_D/L$ in the dimerized phase even
in the presence of a gap.
Specifically, we consider a Hamiltonian of the
form:
\begin{eqnarray}
H= \sum_i [S_i \cdot S_{i+1} + J_2 S_i \cdot S_{i+2} 
+ \delta h (-1)^i S_i \cdot S_{i+1} ]\nonumber\\
\end{eqnarray}
Thus, in correspondence with Eq.~(\ref{GH2}) we have $H_I= (-1)^i
S_i \cdot S_{i+1}$ and we choose the driving coupling to be $J_2$.
This perturbing Hamiltonian represents a \emph{conjugate field} for the dimer
phase. 
%$\chi_D$ then has the same relationship with the \emph{order parameter
%susceptibility} as $\chi_{\rho}$ has with the spin stiffness ($\rho$). 
The scaling dimension of $H_I$ is known~\cite{Affleck_Logarithmic_Corrections}, $\Delta_D=\frac{1}{2}$,
and from Eq.~(\ref{Chi_Scaling}) we therefore find:
\begin{equation}
\chi_D\sim L^{4-2\Delta_D}=L^3 \ \ (\mathrm{at}\ J_2^c)
  \label{eq:chid_scaling}
\end{equation}
Due to the presence of the marginal coupling we cannot expect this relation
to hold for $J_2<J_2^c$. However, the marginal coupling changes sign at $J_2^c$
and is therefore absent at $J_2^c$ where Eq.~(\ref{eq:chid_scaling}) should be exact.~\cite{Eggert:1992}
For $J_2 < 0.241167$ it is known\cite{Affleck_Logarithmic_Corrections} that
logarithmic corrections arising from the marginal coupling for the small system sizes considered here
lead to an effective scaling dimension 
$\Delta_D>\frac{1}{2}$.  At $J_2=0$ Affleck and Bonner~\cite{Affleck_Logarithmic_Corrections} estimated
$\Delta_D=0.71$. Hence, using this results at $J_2=0$, we would expect that $\chi_D\sim L^{2.58}$ which
we find is in good agreement with our results at $J_2=0$.
\begin{figure}[t]
\includegraphics[scale=0.35]{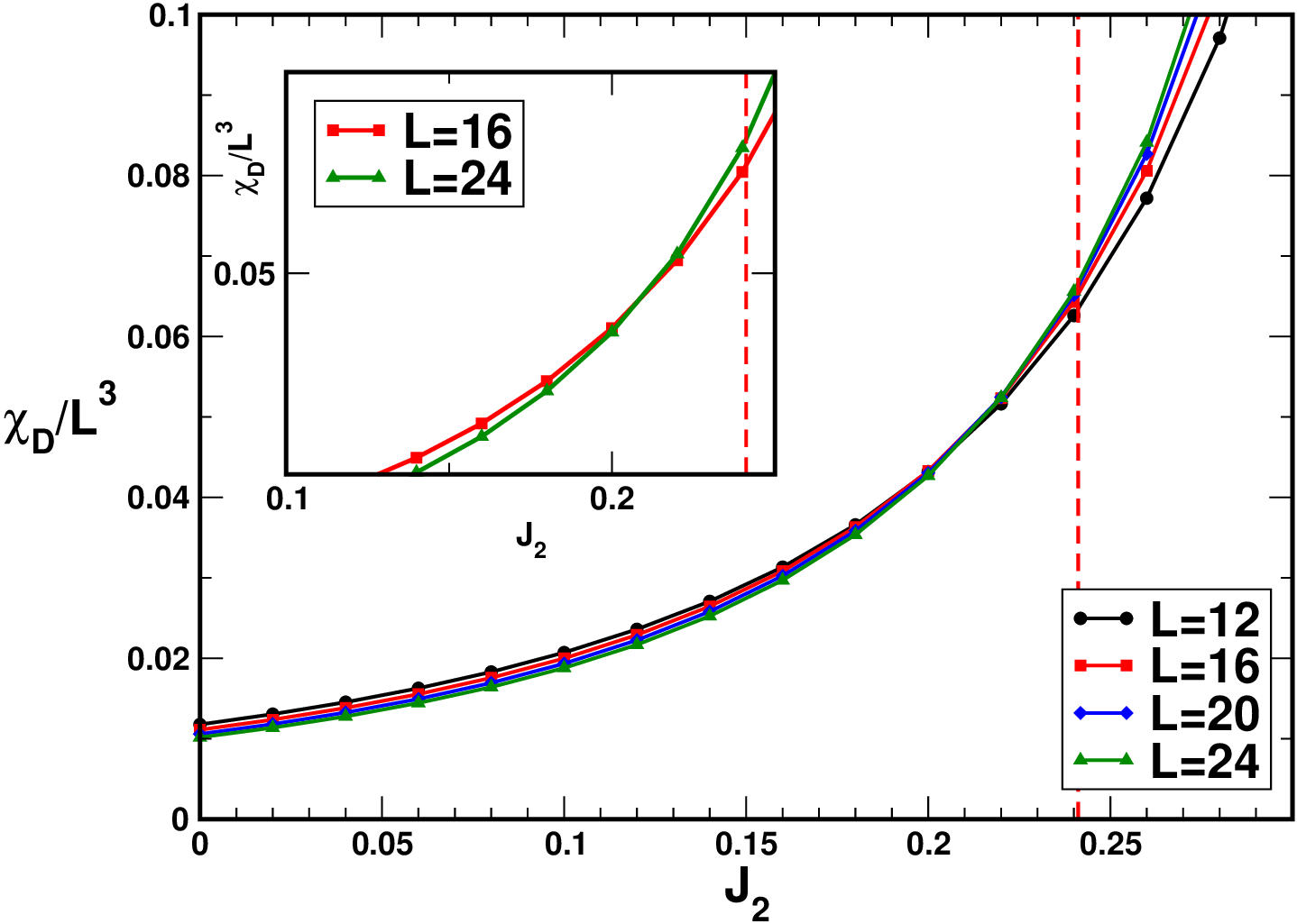}
\caption{\label{Chi_D_vs_L3_and_Inset} (Color online.)  $\frac{\chi_D}{L^3}$ vs. $J_2$.
  The generalized dimer fidelity susceptibility $\chi_D/L^3$ as
    a function of the second nearest-neighbor exchange parameter $J_2$. A
    clear intersection of all curves can be seen in the vicinity of the
    proposed critical point at $J_2\sim 0.2-0.25$.  The inset explicitly shows the
    crossing of $L=12$ and $L=24$. The dashed vertical lines indicate $J_2^c$.} \end{figure}

We now need to consider the case $J_2>0.241167$. At $J_2=1/2$ the model is exactly solvable~\cite{MG} and
the two dimerized ground-states are exactly degenerate even for finite $L$. For $J_2^c<J_2<1/2$ the system
is gapped with a unique ground-state but with an exponentially low-lying excited state. In the thermodynamic limit the
two-fold degeneracy of ground-state is recovered, corresponding to the degeneracy of the two dimerization patterns.
From this it follows that $\chi_D$ is formally infinite at $J_2=0$ and
as $L \rightarrow \infty$ for $J_2^c<J_2<1/2$ we expect $\chi_D$ to diverge exponentially with $L$.
At $J_2^c$ we expect $\chi_D$ to exactly scale as $L^3$ and for $J_2<J_2^c$ we expect $\chi_D\sim L^\alpha_{\mathrm{eff}}$
with $\alpha_{\mathrm{eff}}<3$. Hence, if $\chi_D/L^3$ is plotted for different $L$ we would expect the
curves to cross at $J_2^c$. However, the crossing might be difficult to observe since it effectively
arise from logarithmic corrections.

\begin{figure}[t]
\includegraphics[scale=0.35]{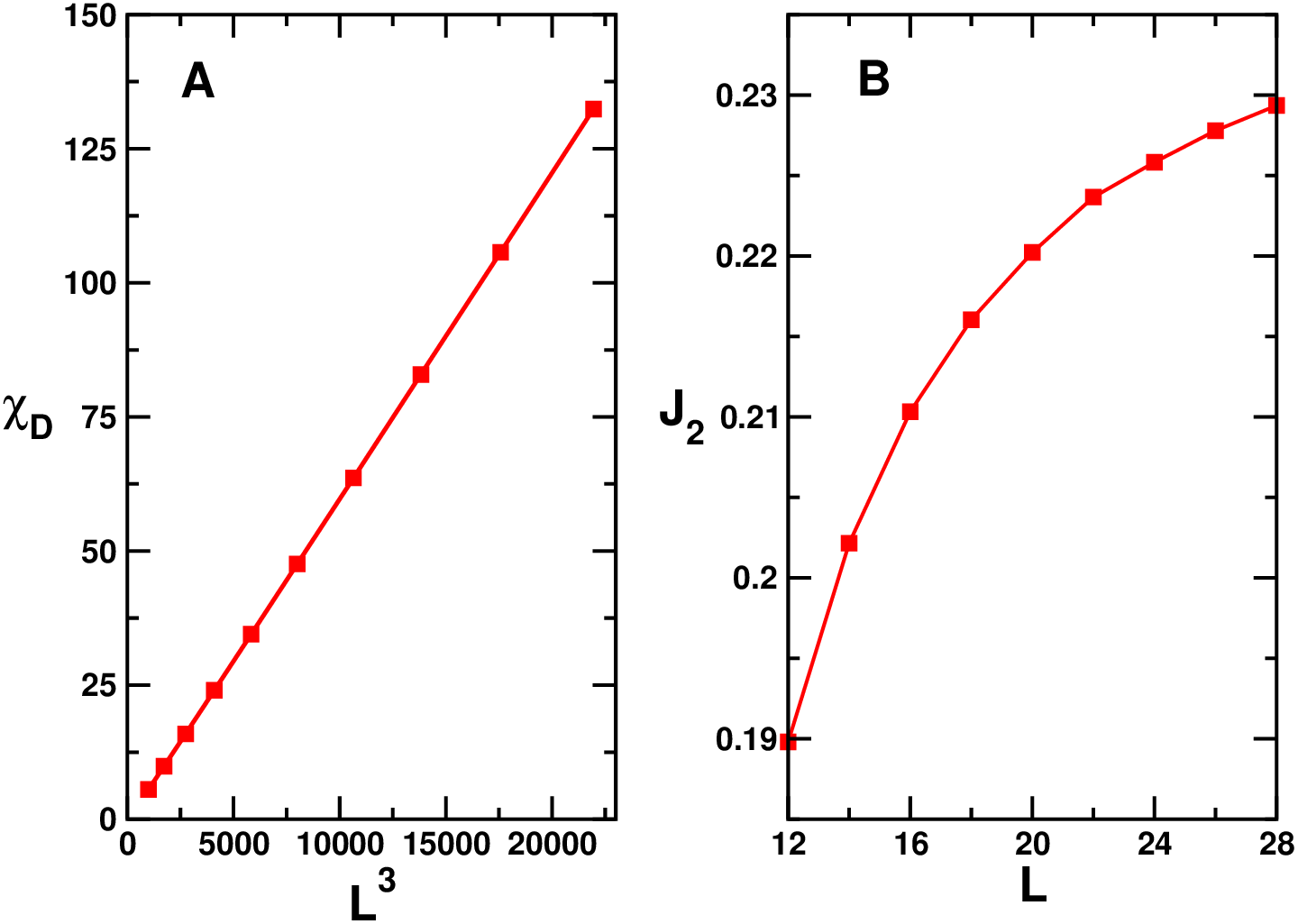}
\caption{\label{Chi_D_Scaling_and_Power_Law}  (Color online.)
  A: Scaling of $\chi_D$ vs. $L^3$ at the critical point $J_2=0.241167$.  A
    near perfectly linear scaling is observed.   B: The $J_2$ value of the
    intersection of $\frac{\chi_D}{L^3}$ between systems of size $L$ and $L+2$ 
        plotted as a function of $L$.  The curve can be fitted
    with a power-law line of best fit.  The line of best fit is found to
    converge to $J_2=0.241$.} \end{figure}

Our results for $\chi_D/L^3$ are shown in 
Fig.~\ref{Chi_D_vs_L3_and_Inset}, where a crossing of the curves are visible
around $J_2\sim 0.2-0.25$. As an illustration, the inset of Fig.~\ref{Chi_D_vs_L3_and_Inset} shows
the crossing of $L=12$ and $L=24$.  In order to obtain a more precise estimate of $J_2^c$
the intersection of each curve and the
curve corresponding to the next largest system were tabulated ($L$ and $L+2$).
These intersection points as a function of system size were then plotted Fig.~\ref{Chi_D_Scaling_and_Power_Law}A
and found to obey a power-law of the form $a-bL^{-\alpha}$ with $\alpha\sim 1.8$ and $a=0.241$.
This estimate of the critical coupling  is in good  agreement with the
value of $J_2^c=0.241167$.~\cite{Eggert}

To further verify the scaling of $\chi_D$ at $J_2^c$ we show in Fig.~\ref{Chi_D_Scaling_and_Power_Law}B
$\chi_D$ at 
$J_2^c$ as a function of the cubed system size, $L^3$.  The
strong linear scaling is in contrast to the scaling a small distance away
from the critical point (not shown) where the scaling was found to be distorted
by logarithmic corrections.  

\section{The AF Fidelity Susceptibility, $\chi_\mathrm{AF}$}\label{sec:chiAF}
Finally, we briefly discuss another fidelity susceptibility very analogous to $\chi_D$.
We consider a perturbing term in the form of a staggered field of the form $\sum_i(-1)^iS^z_i$
with an associated fidelity susceptibility, $\chi_\mathrm{AF}$.
The scaling dimension of such a staggered field is $\Delta_{\mathrm{AF}}=\frac{1}{2}$ and as for $\chi_D$
we therefore expect that $\chi_{\mathrm{AF}}\sim L^3$ at $J_2^c$. However, 
in this case it is known~\cite{Affleck_Logarithmic_Corrections} that the effective scaling dimension
for $J_2<J_2^c$ is {\it smaller} than $\frac{1}{2}$ resulting in $\chi_{\mathrm{AF}}\sim L^{\alpha_{\mathrm{eff}}}$
with $\alpha_{\mathrm{eff}}>3$ for $J_2<J_2^c$. On the other had, in the dimerized phase $\chi_\mathrm{AF}$ must clearly
go to zero exponentially with $L$. Hence, if $\chi_\mathrm{AF}$ is plotted for different $L$ as a function
of $J_2$ a crossing of the curves should occur.
\begin{figure}[t]
\includegraphics[scale=0.35]{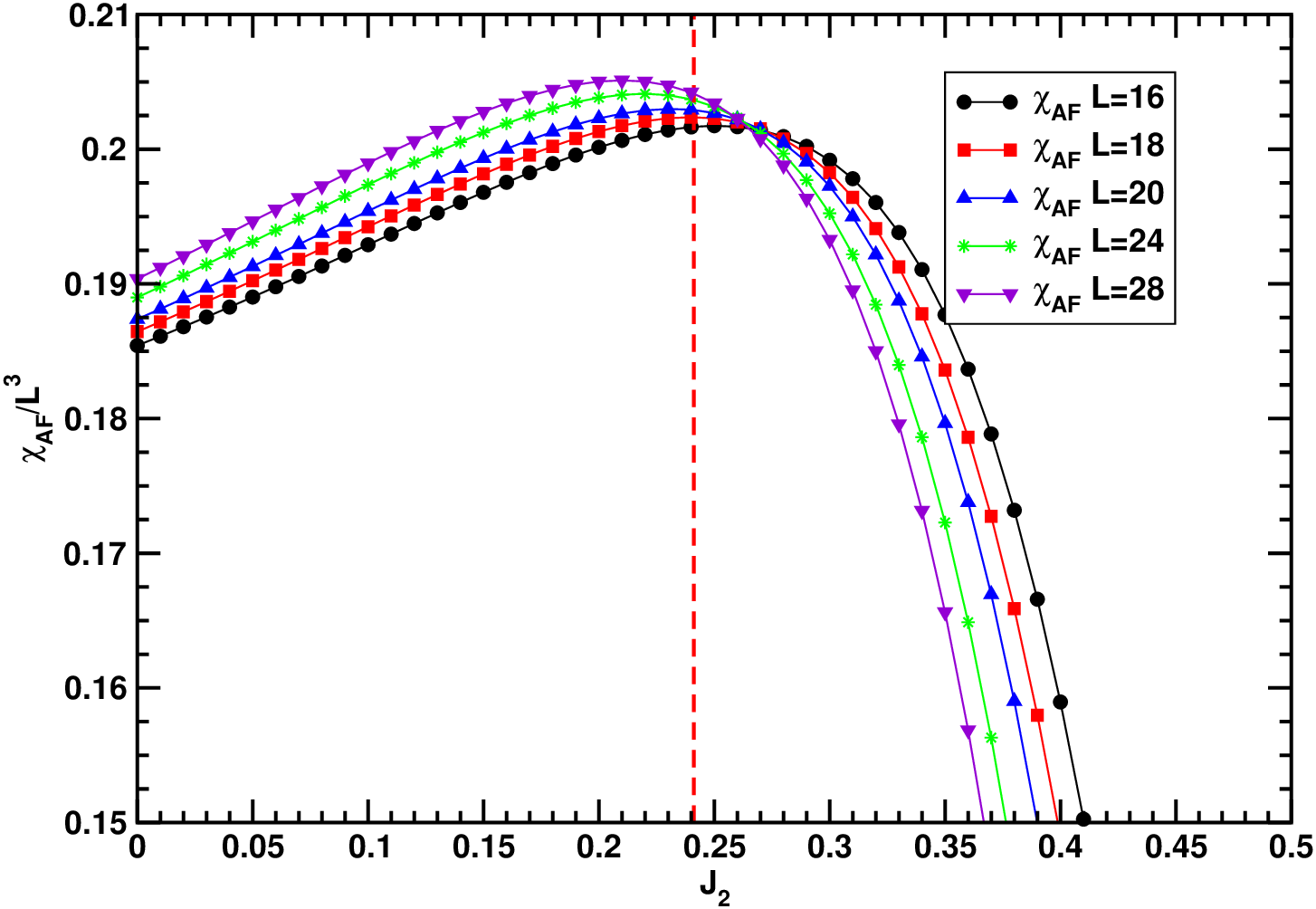}
\caption{\label{ChiAF} (Color online.) $\chi_{\mathrm{AF}}/L^3$ versus $J_2$.
  $\chi_{\mathrm{AF}}$ is expected to approach zero exponentially with the system size for $J_2>J_2^c$,
    to scale as $L^3$ at $J_2^c$ and to scale as $L^{\alpha_\mathrm{eff}}$ with $\alpha_\mathrm{eff}>3$ for
      $J_2<J_2^c$.
  A crossing close to the critical point $J_2^c$ (dashed vertical line) is then visible.}
\end{figure}

Our results are shown in Fig.~\ref{ChiAF} where $\chi_\mathrm{AF}/L^3$ is plotted versus $J_2$ for a
number of system sizes. It is clear from these results that $\chi_\mathrm{AF}$ indeed goes to zero rapidly 
in the dimerized phase as one would expect. Close to $J_2^c$ the scaling is close to $L^3$ where as for $J_2<J_2^c$
it is faster than $L^3$. Hence, as can be seen in Fig.~\ref{ChiAF}, a crossing occurs close to $J_2^c$.

\section{Conclusion and Summary}

In this paper we have demonstrated the potential benefits of extending the
concept of a fidelity susceptibility beyond a simple perturbation of
the same term that drives the quantum phase transition.
By using the spin-$1/2$ Heisenberg spin chain as an
example we first created a susceptibility which was directly coupled to the
spin stiffness but of increased sensitivity.  This
fidelity susceptibility, which we labelled $\chi_{\rho}$ can be used to  successfully
estimate the transition point at $J_2 \sim 0.241$.  Next we constructed
another fidelity susceptibility, $\chi_D$, this time coupled to the order
parameter susceptibility of the dimer phase.  Again, we were able to estimate
the critical point at a value of 0.241.  Finally, we discussed an anti ferromagnetic fidelity susceptibility
that rapidly approaches zero in the dimerized phase but diverges in the Heisenberg phase. Although susceptibilities linked to
these quantities appeared the most useful
for the $J_1-J_2$ model we considered here, it is possible to define many other fidelity susceptibilities that could provide
valuable insights into the ordering occurring in the system being studied.

\begin{acknowledgments}

MT acknowledge many fruitful discussions with Sedigh Ghamari. 
ESS would like to thank Fabien Alet for several discussions about the fidelity susceptibility and
I Affleck for helpful discussions concerning scaling dimensions.
This work was supported by NSERC and by the Shared Hierarchical Academic
  Research Computing Network.
\end{acknowledgments}

\bibliography{References}

\begin{thebibliography}{33}%
\makeatletter
\providecommand \@ifxundefined [1]{%
 \@ifx{#1\undefined}
}%
\providecommand \@ifnum [1]{%
 \ifnum #1\expandafter \@firstoftwo
 \else \expandafter \@secondoftwo
 \fi
}%
\providecommand \@ifx [1]{%
 \ifx #1\expandafter \@firstoftwo
 \else \expandafter \@secondoftwo
 \fi
}%
\providecommand \natexlab [1]{#1}%
\providecommand \enquote  [1]{``#1''}%
\providecommand \bibnamefont  [1]{#1}%
\providecommand \bibfnamefont [1]{#1}%
\providecommand \citenamefont [1]{#1}%
\providecommand \href@noop [0]{\@secondoftwo}%
\providecommand \href [0]{\begingroup \@sanitize@url \@href}%
\providecommand \@href[1]{\@@startlink{#1}\@@href}%
\providecommand \@@href[1]{\endgroup#1\@@endlink}%
\providecommand \@sanitize@url [0]{\catcode `\\12\catcode `\$12\catcode
  `\&12\catcode `\#12\catcode `\^12\catcode `\_12\catcode `\%12\relax}%
\providecommand \@@startlink[1]{}%
\providecommand \@@endlink[0]{}%
\providecommand \url  [0]{\begingroup\@sanitize@url \@url }%
\providecommand \@url [1]{\endgroup\@href {#1}{\urlprefix }}%
\providecommand \urlprefix  [0]{URL }%
\providecommand \Eprint [0]{\href }%
\providecommand \doibase [0]{http://dx.doi.org/}%
\providecommand \selectlanguage [0]{\@gobble}%
\providecommand \bibinfo  [0]{\@secondoftwo}%
\providecommand \bibfield  [0]{\@secondoftwo}%
\providecommand \translation [1]{[#1]}%
\providecommand \BibitemOpen [0]{}%
\providecommand \bibitemStop [0]{}%
\providecommand \bibitemNoStop [0]{.\EOS\space}%
\providecommand \EOS [0]{\spacefactor3000\relax}%
\providecommand \BibitemShut  [1]{\csname bibitem#1\endcsname}%
\let\auto@bib@innerbib\@empty
%</preamble>
\bibitem [{\citenamefont {Sachdev}(1999)}]{Sachdev_QPT_Book}%
  \BibitemOpen
  \bibfield  {author} {\bibinfo {author} {\bibfnamefont {S.}~\bibnamefont
  {Sachdev}},\ }\href@noop {} {\emph {\bibinfo {title} {Quantum Phase
  Transitions}}}\ (\bibinfo  {publisher} {Cambridge University Press},\
  \bibinfo {year} {1999})\BibitemShut {NoStop}%
\bibitem [{\citenamefont {Quan}\ \emph {et~al.}(2006)\citenamefont {Quan},
  \citenamefont {Song}, \citenamefont {Liu}, \citenamefont {Zanardi},\ and\
  \citenamefont {Sun}}]{Quan_Original}%
  \BibitemOpen
  \bibfield  {author} {\bibinfo {author} {\bibfnamefont {H.~T.}\ \bibnamefont
  {Quan}}, \bibinfo {author} {\bibfnamefont {Z.}~\bibnamefont {Song}}, \bibinfo
  {author} {\bibfnamefont {X.~F.}\ \bibnamefont {Liu}}, \bibinfo {author}
  {\bibfnamefont {P.}~\bibnamefont {Zanardi}}, \ and\ \bibinfo {author}
  {\bibfnamefont {C.~P.}\ \bibnamefont {Sun}},\ }\href {\doibase
  10.1103/PhysRevLett.96.140604} {\bibfield  {journal} {\bibinfo  {journal}
  {Phys. Rev. Lett.}\ }\textbf {\bibinfo {volume} {96}},\ \bibinfo {pages}
  {140604} (\bibinfo {year} {2006})}\BibitemShut {NoStop}%
\bibitem [{\citenamefont {Zanardi}\ and\ \citenamefont
  {Paunkovi\ifmmode~\acute{c}\else \'{c}\fi{}}(2006)}]{Zanardi_Original}%
  \BibitemOpen
  \bibfield  {author} {\bibinfo {author} {\bibfnamefont {P.}~\bibnamefont
  {Zanardi}}\ and\ \bibinfo {author} {\bibfnamefont {N.}~\bibnamefont
  {Paunkovi\ifmmode~\acute{c}\else \'{c}\fi{}}},\ }\href {\doibase
  10.1103/PhysRevE.74.031123} {\bibfield  {journal} {\bibinfo  {journal} {Phys.
  Rev. E}\ }\textbf {\bibinfo {volume} {74}},\ \bibinfo {pages} {031123}
  (\bibinfo {year} {2006})}\BibitemShut {NoStop}%
\bibitem [{\citenamefont {Zhou}\ and\ \citenamefont
  {Barjaktarevič}(2008)}]{Zhou_Original}%
  \BibitemOpen
  \bibfield  {author} {\bibinfo {author} {\bibfnamefont {H.-Q.}\ \bibnamefont
  {Zhou}}\ and\ \bibinfo {author} {\bibfnamefont {J.~P.}\ \bibnamefont
  {Barjaktarevič}},\ }\href {http://stacks.iop.org/1751-8121/41/i=41/a=412001}
  {\bibfield  {journal} {\bibinfo  {journal} {Journal of Physics A:
  Mathematical and Theoretical}\ }\textbf {\bibinfo {volume} {41}},\ \bibinfo
  {pages} {412001} (\bibinfo {year} {2008})}\BibitemShut {NoStop}%
\bibitem [{\citenamefont {You}\ \emph {et~al.}(2007)\citenamefont {You},
  \citenamefont {Li},\ and\ \citenamefont {Gu}}]{You}%
  \BibitemOpen
  \bibfield  {author} {\bibinfo {author} {\bibfnamefont {W.-L.}\ \bibnamefont
  {You}}, \bibinfo {author} {\bibfnamefont {Y.-W.}\ \bibnamefont {Li}}, \ and\
  \bibinfo {author} {\bibfnamefont {S.-J.}\ \bibnamefont {Gu}},\ }\href
  {\doibase 10.1103/PhysRevE.76.022101} {\bibfield  {journal} {\bibinfo
  {journal} {Phys. Rev. E}\ }\textbf {\bibinfo {volume} {76}},\ \bibinfo
  {pages} {022101} (\bibinfo {year} {2007})}\BibitemShut {NoStop}%
\bibitem [{\citenamefont {Zanardi}\ \emph
  {et~al.}(2007{\natexlab{a}})\citenamefont {Zanardi}, \citenamefont
  {Cozzini},\ and\ \citenamefont {Giorda}}]{Zanardi_Free_Fermion_Systems}%
  \BibitemOpen
  \bibfield  {author} {\bibinfo {author} {\bibfnamefont {P.}~\bibnamefont
  {Zanardi}}, \bibinfo {author} {\bibfnamefont {M.}~\bibnamefont {Cozzini}}, \
  and\ \bibinfo {author} {\bibfnamefont {P.}~\bibnamefont {Giorda}},\ }\href
  {http://stacks.iop.org/1742-5468/2007/i=02/a=L02002} {\bibfield  {journal}
  {\bibinfo  {journal} {Journal of Statistical Mechanics: Theory and
  Experiment}\ }\textbf {\bibinfo {volume} {2007}},\ \bibinfo {pages} {L02002}
  (\bibinfo {year} {2007}{\natexlab{a}})}\BibitemShut {NoStop}%
\bibitem [{\citenamefont {Cozzini}\ \emph
  {et~al.}(2007{\natexlab{a}})\citenamefont {Cozzini}, \citenamefont {Giorda},\
  and\ \citenamefont {Zanardi}}]{Cozzini_Graphs}%
  \BibitemOpen
  \bibfield  {author} {\bibinfo {author} {\bibfnamefont {M.}~\bibnamefont
  {Cozzini}}, \bibinfo {author} {\bibfnamefont {P.}~\bibnamefont {Giorda}}, \
  and\ \bibinfo {author} {\bibfnamefont {P.}~\bibnamefont {Zanardi}},\ }\href
  {\doibase 10.1103/PhysRevB.75.014439} {\bibfield  {journal} {\bibinfo
  {journal} {Phys. Rev. B}\ }\textbf {\bibinfo {volume} {75}},\ \bibinfo
  {pages} {014439} (\bibinfo {year} {2007}{\natexlab{a}})}\BibitemShut
  {NoStop}%
\bibitem [{\citenamefont {Cozzini}\ \emph
  {et~al.}(2007{\natexlab{b}})\citenamefont {Cozzini}, \citenamefont
  {Ionicioiu},\ and\ \citenamefont {Zanardi}}]{Cozzini_Matrix_Product_States}%
  \BibitemOpen
  \bibfield  {author} {\bibinfo {author} {\bibfnamefont {M.}~\bibnamefont
  {Cozzini}}, \bibinfo {author} {\bibfnamefont {R.}~\bibnamefont {Ionicioiu}},
  \ and\ \bibinfo {author} {\bibfnamefont {P.}~\bibnamefont {Zanardi}},\ }\href
  {\doibase 10.1103/PhysRevB.76.104420} {\bibfield  {journal} {\bibinfo
  {journal} {Phys. Rev. B}\ }\textbf {\bibinfo {volume} {76}},\ \bibinfo
  {pages} {104420} (\bibinfo {year} {2007}{\natexlab{b}})}\BibitemShut
  {NoStop}%
\bibitem [{\citenamefont {Buonsante}\ and\ \citenamefont
  {Vezzani}(2007)}]{Buonsante_Bose_Hubbard}%
  \BibitemOpen
  \bibfield  {author} {\bibinfo {author} {\bibfnamefont {P.}~\bibnamefont
  {Buonsante}}\ and\ \bibinfo {author} {\bibfnamefont {A.}~\bibnamefont
  {Vezzani}},\ }\href {\doibase 10.1103/PhysRevLett.98.110601} {\bibfield
  {journal} {\bibinfo  {journal} {Phys. Rev. Lett.}\ }\textbf {\bibinfo
  {volume} {98}},\ \bibinfo {pages} {110601} (\bibinfo {year}
  {2007})}\BibitemShut {NoStop}%
\bibitem [{\citenamefont {Schwandt}\ \emph {et~al.}(2009)\citenamefont
  {Schwandt}, \citenamefont {Alet},\ and\ \citenamefont
  {Capponi}}]{Schwandt:2009}%
  \BibitemOpen
  \bibfield  {author} {\bibinfo {author} {\bibfnamefont {D.}~\bibnamefont
  {Schwandt}}, \bibinfo {author} {\bibfnamefont {F.}~\bibnamefont {Alet}}, \
  and\ \bibinfo {author} {\bibfnamefont {S.}~\bibnamefont {Capponi}},\
  }\href@noop {} {\bibfield  {journal} {\bibinfo  {journal} {Physical Review
  Letters}\ }\textbf {\bibinfo {volume} {103}},\ \bibinfo {pages} {170501}
  (\bibinfo {year} {2009})}\BibitemShut {NoStop}%
\bibitem [{\citenamefont {Albuquerque}\ \emph {et~al.}(2010)\citenamefont
  {Albuquerque}, \citenamefont {Alet}, \citenamefont {Sire},\ and\
  \citenamefont {Capponi}}]{Albuquerque:2010}%
  \BibitemOpen
  \bibfield  {author} {\bibinfo {author} {\bibfnamefont {A.~F.}\ \bibnamefont
  {Albuquerque}}, \bibinfo {author} {\bibfnamefont {F.}~\bibnamefont {Alet}},
  \bibinfo {author} {\bibfnamefont {C.}~\bibnamefont {Sire}}, \ and\ \bibinfo
  {author} {\bibfnamefont {S.}~\bibnamefont {Capponi}},\ }\href {\doibase
  10.1103/PhysRevB.81.064418} {\bibfield  {journal} {\bibinfo  {journal}
  {Physical Review B}\ }\textbf {\bibinfo {volume} {81}},\ \bibinfo {pages}
  {064418} (\bibinfo {year} {2010})}\BibitemShut {NoStop}%
\bibitem [{\citenamefont {Chen}\ \emph {et~al.}(2007)\citenamefont {Chen},
  \citenamefont {Wang}, \citenamefont {Gu},\ and\ \citenamefont
  {Wang}}]{Chen_J1_J2}%
  \BibitemOpen
  \bibfield  {author} {\bibinfo {author} {\bibfnamefont {S.}~\bibnamefont
  {Chen}}, \bibinfo {author} {\bibfnamefont {L.}~\bibnamefont {Wang}}, \bibinfo
  {author} {\bibfnamefont {S.-J.}\ \bibnamefont {Gu}}, \ and\ \bibinfo {author}
  {\bibfnamefont {Y.}~\bibnamefont {Wang}},\ }\href {\doibase
  10.1103/PhysRevE.76.061108} {\bibfield  {journal} {\bibinfo  {journal} {Phys.
  Rev. E}\ }\textbf {\bibinfo {volume} {76}},\ \bibinfo {pages} {061108}
  (\bibinfo {year} {2007})}\BibitemShut {NoStop}%
\bibitem [{\citenamefont {{Gu}}(2010)}]{Gu_Review}%
  \BibitemOpen
  \bibfield  {author} {\bibinfo {author} {\bibfnamefont {S.}~\bibnamefont
  {{Gu}}},\ }\href {\doibase 10.1142/S0217979210056335} {\bibfield  {journal}
  {\bibinfo  {journal} {International Journal of Modern Physics B}\ }\textbf
  {\bibinfo {volume} {24}},\ \bibinfo {pages} {4371} (\bibinfo {year}
  {2010})}\BibitemShut {NoStop}%
\bibitem [{\citenamefont {Zanardi}\ \emph
  {et~al.}(2007{\natexlab{b}})\citenamefont {Zanardi}, \citenamefont {Giorda},\
  and\ \citenamefont {Cozzini}}]{Zanardi:2007}%
  \BibitemOpen
  \bibfield  {author} {\bibinfo {author} {\bibfnamefont {P.}~\bibnamefont
  {Zanardi}}, \bibinfo {author} {\bibfnamefont {P.}~\bibnamefont {Giorda}}, \
  and\ \bibinfo {author} {\bibfnamefont {M.}~\bibnamefont {Cozzini}},\
  }\href@noop {} {\bibfield  {journal} {\bibinfo  {journal} {Physical Review
  Letters}\ }\textbf {\bibinfo {volume} {99}},\ \bibinfo {pages} {100603}
  (\bibinfo {year} {2007}{\natexlab{b}})}\BibitemShut {NoStop}%
\bibitem [{\citenamefont {Campos~Venuti}\ and\ \citenamefont
  {Zanardi}(2007)}]{Venuti_Scaling}%
  \BibitemOpen
  \bibfield  {author} {\bibinfo {author} {\bibfnamefont {L.}~\bibnamefont
  {Campos~Venuti}}\ and\ \bibinfo {author} {\bibfnamefont {P.}~\bibnamefont
  {Zanardi}},\ }\href {\doibase 10.1103/PhysRevLett.99.095701} {\bibfield
  {journal} {\bibinfo  {journal} {Phys. Rev. Lett.}\ }\textbf {\bibinfo
  {volume} {99}},\ \bibinfo {pages} {095701} (\bibinfo {year}
  {2007})}\BibitemShut {NoStop}%
\bibitem [{\citenamefont {De~Grandi}\ \emph {et~al.}(2011)\citenamefont
  {De~Grandi}, \citenamefont {Polkovnikov},\ and\ \citenamefont
  {Sandvik}}]{DeGrandi:2011}%
  \BibitemOpen
  \bibfield  {author} {\bibinfo {author} {\bibfnamefont {C.}~\bibnamefont
  {De~Grandi}}, \bibinfo {author} {\bibfnamefont {A.}~\bibnamefont
  {Polkovnikov}}, \ and\ \bibinfo {author} {\bibfnamefont {A.~W.}\ \bibnamefont
  {Sandvik}},\ }\href {http://arxiv.org/abs/1106.4078v2} {\bibfield  {journal}
  {\bibinfo  {journal} {arXiv.org}\ }\textbf {\bibinfo {volume}
  {cond-mat.other}} (\bibinfo {year} {2011})}\BibitemShut {NoStop}%
\bibitem [{\citenamefont {Gu}\ and\ \citenamefont {Lin}(2009)}]{Gu:2009}%
  \BibitemOpen
  \bibfield  {author} {\bibinfo {author} {\bibfnamefont {S.}~\bibnamefont
  {Gu}}\ and\ \bibinfo {author} {\bibfnamefont {H.}~\bibnamefont {Lin}},\
  }\href@noop {} {\bibfield  {journal} {\bibinfo  {journal} {EPL (Europhysics
  Letters)}\ }\textbf {\bibinfo {volume} {87}},\ \bibinfo {pages} {10003}
  (\bibinfo {year} {2009})}\BibitemShut {NoStop}%
\bibitem [{\citenamefont {Chen}\ \emph {et~al.}(2008)\citenamefont {Chen},
  \citenamefont {Wang}, \citenamefont {Hao},\ and\ \citenamefont
  {Wang}}]{Chen_Perturbation}%
  \BibitemOpen
  \bibfield  {author} {\bibinfo {author} {\bibfnamefont {S.}~\bibnamefont
  {Chen}}, \bibinfo {author} {\bibfnamefont {L.}~\bibnamefont {Wang}}, \bibinfo
  {author} {\bibfnamefont {Y.}~\bibnamefont {Hao}}, \ and\ \bibinfo {author}
  {\bibfnamefont {Y.}~\bibnamefont {Wang}},\ }\href {\doibase
  10.1103/PhysRevA.77.032111} {\bibfield  {journal} {\bibinfo  {journal} {Phys.
  Rev. A}\ }\textbf {\bibinfo {volume} {77}},\ \bibinfo {pages} {032111}
  (\bibinfo {year} {2008})}\BibitemShut {NoStop}%
\bibitem [{\citenamefont {Haldane}(1982)}]{J1_J2_Field_Theory_1}%
  \BibitemOpen
  \bibfield  {author} {\bibinfo {author} {\bibfnamefont {F.~D.~M.}\
  \bibnamefont {Haldane}},\ }\href {\doibase 10.1103/PhysRevB.25.4925}
  {\bibfield  {journal} {\bibinfo  {journal} {Phys. Rev. B}\ }\textbf {\bibinfo
  {volume} {25}},\ \bibinfo {pages} {4925} (\bibinfo {year}
  {1982})}\BibitemShut {NoStop}%
\bibitem [{\citenamefont {Eggert}\ and\ \citenamefont
  {Affleck}(1992)}]{Eggert:1992}%
  \BibitemOpen
  \bibfield  {author} {\bibinfo {author} {\bibfnamefont {S.}~\bibnamefont
  {Eggert}}\ and\ \bibinfo {author} {\bibfnamefont {I.}~\bibnamefont
  {Affleck}},\ }\href@noop {} {\bibfield  {journal} {\bibinfo  {journal}
  {Physical Review B}\ }\textbf {\bibinfo {volume} {46}},\ \bibinfo {pages}
  {10866} (\bibinfo {year} {1992})}\BibitemShut {NoStop}%
\bibitem [{\citenamefont {Tonegawa}\ and\ \citenamefont
  {Harada}(1987)}]{J1_J2_ED_1}%
  \BibitemOpen
  \bibfield  {author} {\bibinfo {author} {\bibfnamefont {T.}~\bibnamefont
  {Tonegawa}}\ and\ \bibinfo {author} {\bibfnamefont {I.}~\bibnamefont
  {Harada}},\ }\href@noop {} {\bibfield  {journal} {\bibinfo  {journal} {J.
  Phys. Soc. Jpn.}\ }\textbf {\bibinfo {volume} {56}},\ \bibinfo {pages} {2153}
  (\bibinfo {year} {1987})}\BibitemShut {NoStop}%
\bibitem [{\citenamefont {Eggert}(1996)}]{Eggert}%
  \BibitemOpen
  \bibfield  {author} {\bibinfo {author} {\bibfnamefont {S.}~\bibnamefont
  {Eggert}},\ }\href {\doibase 10.1103/PhysRevB.54.R9612} {\bibfield  {journal}
  {\bibinfo  {journal} {Phys. Rev. B}\ }\textbf {\bibinfo {volume} {54}},\
  \bibinfo {pages} {R9612} (\bibinfo {year} {1996})}\BibitemShut {NoStop}%
\bibitem [{\citenamefont {Chitra}\ \emph {et~al.}(1995)\citenamefont {Chitra},
  \citenamefont {Pati}, \citenamefont {Krishnamurthy}, \citenamefont {Sen},\
  and\ \citenamefont {Ramasesha}}]{DMRG_2}%
  \BibitemOpen
  \bibfield  {author} {\bibinfo {author} {\bibfnamefont {R.}~\bibnamefont
  {Chitra}}, \bibinfo {author} {\bibfnamefont {S.}~\bibnamefont {Pati}},
  \bibinfo {author} {\bibfnamefont {H.~R.}\ \bibnamefont {Krishnamurthy}},
  \bibinfo {author} {\bibfnamefont {D.}~\bibnamefont {Sen}}, \ and\ \bibinfo
  {author} {\bibfnamefont {S.}~\bibnamefont {Ramasesha}},\ }\href {\doibase
  10.1103/PhysRevB.52.6581} {\bibfield  {journal} {\bibinfo  {journal} {Phys.
  Rev. B}\ }\textbf {\bibinfo {volume} {52}},\ \bibinfo {pages} {6581}
  (\bibinfo {year} {1995})}\BibitemShut {NoStop}%
\bibitem [{\citenamefont {White}\ and\ \citenamefont {Affleck}(1996)}]{DMRG_3}%
  \BibitemOpen
  \bibfield  {author} {\bibinfo {author} {\bibfnamefont {S.~R.}\ \bibnamefont
  {White}}\ and\ \bibinfo {author} {\bibfnamefont {I.}~\bibnamefont
  {Affleck}},\ }\href {\doibase 10.1103/PhysRevB.54.9862} {\bibfield  {journal}
  {\bibinfo  {journal} {Phys. Rev. B}\ }\textbf {\bibinfo {volume} {54}},\
  \bibinfo {pages} {9862} (\bibinfo {year} {1996})}\BibitemShut {NoStop}%
\bibitem [{\citenamefont {Shastry}\ and\ \citenamefont
  {Sutherland}(1990)}]{Shastry:1990}%
  \BibitemOpen
  \bibfield  {author} {\bibinfo {author} {\bibfnamefont {B.~S.}\ \bibnamefont
  {Shastry}}\ and\ \bibinfo {author} {\bibfnamefont {B.}~\bibnamefont
  {Sutherland}},\ }\href {\doibase 10.1103/PhysRevLett.65.243} {\bibfield
  {journal} {\bibinfo  {journal} {Phys. Rev. Lett.}\ }\textbf {\bibinfo
  {volume} {65}},\ \bibinfo {pages} {243} (\bibinfo {year} {1990})}\BibitemShut
  {NoStop}%
\bibitem [{\citenamefont {Sutherland}\ and\ \citenamefont
  {Shastry}(1990)}]{Sutherland:1990}%
  \BibitemOpen
  \bibfield  {author} {\bibinfo {author} {\bibfnamefont {B.}~\bibnamefont
  {Sutherland}}\ and\ \bibinfo {author} {\bibfnamefont {B.~S.}\ \bibnamefont
  {Shastry}},\ }\href {\doibase 10.1103/PhysRevLett.65.1833} {\bibfield
  {journal} {\bibinfo  {journal} {Phys. Rev. Lett.}\ }\textbf {\bibinfo
  {volume} {65}},\ \bibinfo {pages} {1833} (\bibinfo {year}
  {1990})}\BibitemShut {NoStop}%
\bibitem [{\citenamefont {Laflorencie}\ \emph {et~al.}(2001)\citenamefont
  {Laflorencie}, \citenamefont {Capponi},\ and\ \citenamefont
  {S{\o}rensen}}]{Sorensen_XXZ}%
  \BibitemOpen
  \bibfield  {author} {\bibinfo {author} {\bibfnamefont {N.}~\bibnamefont
  {Laflorencie}}, \bibinfo {author} {\bibfnamefont {S.}~\bibnamefont
  {Capponi}}, \ and\ \bibinfo {author} {\bibfnamefont {E.~S.}\ \bibnamefont
  {S{\o}rensen}},\ }\href@noop {} {\bibfield  {journal} {\bibinfo  {journal}
  {Eur. Phys. J. B}\ }\textbf {\bibinfo {volume} {24}},\ \bibinfo {pages} {77}
  (\bibinfo {year} {2001})}\BibitemShut {NoStop}%
\bibitem [{\citenamefont {Yang}(2007)}]{Yang:2007}%
  \BibitemOpen
  \bibfield  {author} {\bibinfo {author} {\bibfnamefont {M.}~\bibnamefont
  {Yang}},\ }\href@noop {} {\bibfield  {journal} {\bibinfo  {journal} {Physical
  Review B}\ }\textbf {\bibinfo {volume} {76}},\ \bibinfo {pages} {180403}
  (\bibinfo {year} {2007})}\BibitemShut {NoStop}%
\bibitem [{\citenamefont {Fj{\ae}restad}(2008)}]{Fjaerestad:2008}%
  \BibitemOpen
  \bibfield  {author} {\bibinfo {author} {\bibfnamefont {J.}~\bibnamefont
  {Fj{\ae}restad}},\ }\href@noop {} {\bibfield  {journal} {\bibinfo  {journal}
  {Journal of Statistical Mechanics: Theory and Experiment}\ }\textbf {\bibinfo
  {volume} {2008}},\ \bibinfo {pages} {P07011} (\bibinfo {year}
  {2008})}\BibitemShut {NoStop}%
\bibitem [{\citenamefont {Lin}(1990)}]{Lin}%
  \BibitemOpen
  \bibfield  {author} {\bibinfo {author} {\bibfnamefont {H.~Q.}\ \bibnamefont
  {Lin}},\ }\href {\doibase 10.1103/PhysRevB.42.6561} {\bibfield  {journal}
  {\bibinfo  {journal} {Phys. Rev. B}\ }\textbf {\bibinfo {volume} {42}},\
  \bibinfo {pages} {6561} (\bibinfo {year} {1990})}\BibitemShut {NoStop}%
\bibitem [{\citenamefont {Sirker}\ \emph {et~al.}(2011)\citenamefont {Sirker},
  \citenamefont {Pereira},\ and\ \citenamefont {Affleck}}]{Sirker:2011}%
  \BibitemOpen
  \bibfield  {author} {\bibinfo {author} {\bibfnamefont {J.}~\bibnamefont
  {Sirker}}, \bibinfo {author} {\bibfnamefont {R.}~\bibnamefont {Pereira}}, \
  and\ \bibinfo {author} {\bibfnamefont {I.}~\bibnamefont {Affleck}},\ }\href
  {http://link.aps.org/doi/10.1103/PhysRevB.83.035115} {\bibfield  {journal}
  {\bibinfo  {journal} {Physical Review B}\ ,\ \bibinfo {pages} {035115}}
  (\bibinfo {year} {2011})}\BibitemShut {NoStop}%
\bibitem [{\citenamefont {Affleck}\ and\ \citenamefont
  {Bonner}(1990)}]{Affleck_Logarithmic_Corrections}%
  \BibitemOpen
  \bibfield  {author} {\bibinfo {author} {\bibfnamefont {I.}~\bibnamefont
  {Affleck}}\ and\ \bibinfo {author} {\bibfnamefont {J.~C.}\ \bibnamefont
  {Bonner}},\ }\href {\doibase 10.1103/PhysRevB.42.954} {\bibfield  {journal}
  {\bibinfo  {journal} {Phys. Rev. B}\ }\textbf {\bibinfo {volume} {42}},\
  \bibinfo {pages} {954} (\bibinfo {year} {1990})}\BibitemShut {NoStop}%
\bibitem [{\citenamefont {Majumdar}(1970)}]{MG}%
  \BibitemOpen
  \bibfield  {author} {\bibinfo {author} {\bibfnamefont {C.~K.}\ \bibnamefont
  {Majumdar}},\ }\href@noop {} {\bibfield  {journal} {\bibinfo  {journal}
  {Journal of Physics C: Solid State Physics}\ }\textbf {\bibinfo {volume}
  {3}},\ \bibinfo {pages} {911} (\bibinfo {year} {1970})}\BibitemShut {NoStop}%
\end{thebibliography}%

\end{document}